\documentclass[11pt]{article}

\usepackage{jcappub,natbib}
\def\ga{\mathrel{\raise.3ex\hbox{$>$\kern-.75em\lower1ex\hbox{$\sim$}}}}
\def\la{\mathrel{\raise.3ex\hbox{$<$\kern-.75em\lower1ex\hbox{$\sim$}}}}

\def\be{\begin{equation}}
\def\ee{\end{equation}}
\def\beq{\begin{equation}}
\def\eeq{\end{equation}}
\def\beqa{\begin{eqnarray}}
\def\eeqa{\end{eqnarray}}

\usepackage[utf8]{inputenc}
\usepackage{enumerate}
\usepackage{graphicx,amsmath,amsfonts,amssymb,slashed}
\usepackage{bbold,wasysym}

\usepackage{amsmath,amsfonts,bbm,euscript}
\usepackage{enumerate}
\usepackage{xcolor}
\usepackage{color}
\usepackage{hyperref}

\hypersetup{colorlinks, citecolor=blue, linkcolor=red, urlcolor=blue}

\begin{document}

\title{\LARGE {Dyonic Reissner-Nordstrom Black Holes and Superradiant Stability } }
\author[a, b,c]{Yi-Feng Zou,}
\author[d]{Jun-Huai Xu,}
\author[e]{Zhan-Feng Mai,}
\author[a, b, c]{Jia-Hui Huang}

\affiliation[a]{Guangdong-Hong Kong Joint Laboratory of Quantum Matter,
	Southern Nuclear Science Computing Center,
	South China Normal University, Guangzhou 510006, China}
\affiliation[b]{Guangdong Provincial Key Laboratory of Nuclear Science, Institute of quantum matter,
	South China Normal University, Guangzhou 510006, China}
\affiliation[c]{Guangdong Provincial Key Laboratory of Quantum Engineering and Quantum Materials,
	School of Physics and Telecommunication Engineering,
	South China Normal University, Guangzhou 510006, China}
\affiliation[d]{School of Information and Optoelectronic Science and Technology,
	South China Normal University,Guangzhou 510006,China}
\affiliation[e]{Center for Joint Quantum Studies and Department of Physics, School of Science,
	Tianjin University, Yaguan Road 135, Jinnan District, 300350 Tianjin, P. R. China}
\emailAdd{huangjh@m.scnu.edu.cn}

\abstract{\\
Black holes immersed in magnetic fields are believed to be  important systems in  astrophysics. One interesting topic on these systems is their superradiant stability property. In the present paper, we analytically obtain the superradiantly stable regime for the  asymptotically flat dyonic Reissner-Nordstrom black holes with charged massive scalar perturbation.
The effective potential experienced by the scalar perturbation in the dyonic black hole background is obtained and analyzed. It is found that the dyonic black hole is superradiantly stable in the regime $0<r_{-}/r_{+}<2/3$, where $r_\pm$ are the event horizons of the dyonic black hole. Compared with the purely electrically charged Reissner-Nordstrom black hole case, our result indicates that the additional coupling of the charged scalar perturbation with the magnetic field makes the black hole and scalar perturbation system more superradiantly unstable, which provides further evidence on the instability induced by magnetic field in black hole superradiance process.}
\maketitle
\section{Introduction}

Black holes are relativistic objects which are massive and curious in the universe. Aspects of black hole physics have been studied extensively for a long time. Stability is one of the interesting problems in black hole physics. Superradiance is an important factor which affects the stability of the charged or rotating black holes\cite{Brito:2015oca}.
When a (charged) bosonic wave is scattering off a black hole, the bosonic wave can be amplified under certain conditions through extracting energy from the black hole. This is called a superradiance process. For a charged rotating black hole, the superradiance condition is
\begin{equation}
\omega<m \Omega_{H}+q \Phi_{H},
\end{equation}
where $m$ and $q$ are azimuthal quantum number and the charge of the incoming bosonic wave, $\Omega_{H}$ is the angular velocity of the black hole horizon and $\Phi_{H}$ is the electromagnetic potential of the black hole horizon\cite{P1969,Ch1970,M1972,Ya1971,Bardeen1972,Bekenstein1973,Ya1972}. If there is a  mirror between the black hole horizon and spatial infinity, the amplified superradiant modes may be reflected back and forth and grows exponentially, which is the so-called black hole bomb mechanism proposed by Press and Teukolsky\cite{PTbomb}.

Various kinds of black holes have been studied extensively about their superradiant (in)stability in the literature.
Regge and Wheeler proved that the spherically symmetric Schwarzschild black hole is stable under perturbations\cite{wheeler1957}. In the asymptotically flat background, the charged Reissner-Nordstrom(RN) black hole proved superradiantly stable against charged massive perturbation \cite{Hod:2013eea,Hod:2013nn,Huang:2015jza,Hod:2015hza,DiMenza:2014vpa,Xu:2020fgq,Lin:2021ssw,Huang:2021dpa}. The reason is that if the parameters of RN black holes and perturbations satisfy superradiant conditions, there is no effective trapping potential well(mirror) outside the black hole horizon, which reflects the superradiant modes back and forth. A similar case exists for charged RN black holes in string theory. The stringy RN black hole is shown to be  superradiantly stable against charged massive scalar perturbation\cite{Li:2013jna}.

When a mirror or a cavity is placed around a charged RN black hole, it is proved that this black hole is superradiantly unstable under charged massive scalar perturbation in certain parameter spaces \cite{Herdeiro:2013pia,Li:2014gfg,Degollado:2013bha,Sanchis-Gual:2015lje,Fierro:2017fky}. If the charged RN black holes are in asymptotically curved backgrounds, such as anti-de Sitter/de Sitter(AdS/dS) space, these curved backgrounds may provide  natural mirror-like boundary conditions which lead to superradiant instability of the black hole and perturbation systems \cite{Wang:2014eha,Bosch:2016vcp,Huang:2016zoz,Gonzalez:2017shu,Zhu:2014sya}. A similar case exists for charged RN black holes in string theory. When a mirror is introduced, superradiant modes are supported and the stringy RN black hole becomes superradiantly unstable under charged massive scalar perturbation \cite{Li:2014xxa,Li:2014fna,Li:2015mqa}. It is also found that extra coupling between the scalar field and the gravity can result in superradiant instability of RN/RN-AdS black holes \cite{Kolyvaris:2018zxl,Abdalla:2019irr}.

In addition, previous studies imply that magnetic field surrounding a black hole can also provide a confining mechanism. In the weak magnetic field approximation, Refs.\cite{Konoplya:2007yy,Konoplya:2008hj}  showed that when a scalar field is propagating on the Ernst background\cite{Ernst:1976mzr}, the magnetic field can induce an effective mass $\mu_{\text{eff}}\varpropto B$ ($B$ is the magnetic field strength) for the scalar field. A first fully-consistent linear analysis of
the superradiant instability of the Ernst spacetime and scalar perturbations is given in \cite{Brito:2014nja}. By studying scalar perturbation of a magnetized Kerr–Newman
black hole, the authors confirm  the details of the superradiant instability and find a constraint on the black-hole spin and the surrounding magnetic field.

The influence of magnetic fields on black hole superradiance is an interesting topic and may have astrophysical application. The Ernst black hole is not asymptotically flat and describes a black hole immersed in an asymptotically uniform magnetic field.  In this paper we will discuss the superradiant stability of a class of asymptotically flat magnetically charged black holes-dyonic RN black holes\cite{Bin:2013}. The spacetime metric of this kind of black holes are very similar with RN black holes, however, there is an additional magnetic field. By comparing the superradiant stability properties of the dyonic RN black holes with that of the RN black holes, we can clearly see the effect of the magnetic field on black hole superradiance.

The paper is organized as follows. In Section $2$, we describe the dyonic RN black hole and scalar perturbation system and analyze the angular part and radial part of the equation of motion of the scalar perturbation. In Section $3$, we derive the effective potential experienced by the scalar perturbation and analyze the asymptotical behaviors of the effective potential. In Section $4$, we carefully analyze the shape of the effective potential and get the superradiantly stable parameter region for the system. Section $5$ is devoted to the summary.

\section{Equations of Motion of the Scalar Perturbation}
\label{sec:spinsmfp}

The dyonic RN black hole is a stationary and spherically symmetric spacetime geometry, which is a solution of the Einstein-Maxwell theory \cite{Bin:2013}. Using the spherical coordinates($t,r,\theta,\phi$), the line element can be expressed in the form ( we use natural unit in which $G=c=\hbar$=1)
\begin{equation}
	ds^{2}=-\frac{ \Box}{r^2}\text{dt}^2+\frac{ r^2}{\Box}\text{dr}^2+ r^2\text{d$\theta $}^2+ r^2 \sin ^2\theta \text{d$\phi $}^2,
\end{equation}
where
\begin{equation}
\Box=-2 M r+r^{2}+Q_e^2+Q_m^2,
\end{equation}
$M$ is the mass of the black hole, $Q_{e}$ and $Q_{m}$ are electric and magnetic charges of the black hole respectively.
 The dyonic RN black hole  has an outer horizon at $r_{+}$ and an inner horizon at $r_{-}$,
\begin{equation}
	r_{+}=M +\sqrt{M^2-Q_e^2-Q_m^2},\quad 	r_{-}=M -\sqrt{M^2-Q_e^2-Q_m^2}.
\end{equation}
Obviously, they satisfy the following  relations
\begin{equation}
	\Box=(r-r_{+})(r-r_{-}) ,\quad r_{+}r_{-}=Q_e^2+Q_m^2,\quad r_{+}+r_{-}=2M.
	\label{key18}
\end{equation}

The equation of motion for an electrically charged massive scalar perturbation $\Phi$ in the dyonic RN black hole background is descried by the covariant Klein-Gordon (KG) equation
\begin{equation}
	(D^{\nu}D_{\nu}-\mu^{2})\Phi=0,
\end{equation}
where $D^{\nu}=\bigtriangledown^{\nu}-i q A^{\nu}$ and $D_{\nu}$=$\bigtriangledown_{\nu}-i q A_{\nu}$  are the covariant derivatives, $q$ and $\mu$ are the charge and mass of the scalar field respectively. The electromagnetic field of the dyonic black hole is described by the following vector potential
\begin{equation}
	A_{\nu}=(-\frac{Q_{e}}{r},0,0,	Q_{m}(\cos \theta\mp1)),
\end{equation}
where the upper minus sign applies to the north half-sphere of the black hole and the lower plus sign applies to the south half-sphere\cite{Bin:2013}.

 The  solution of the KG equation can be decomposed as following form
\begin{equation}
	\Phi(t,r,\theta,\phi)=R( r ) Y( \theta) e^{im\phi}e^{-i\omega t},
	\label{key6}
\end{equation}
where $\omega$ is the angular frequency of the scalar perturbation and $m$ is azimuthal harmonic index.
 $Y( \theta)$  is the angular part  and $R( r )$ is the radial part of the solution. Plugging the above solution into the KG equation, we can get the radial and angular parts of the equation of motion.
 Considering  the electromagnetic potentials of the northern and southern hemispheres are different, we will discuss  the angular equation of motion
 in two cases in the following.
\subsection{The North Half-Sphere of the Black Hole($0\leqslant\theta\leqslant\frac{\pi}{2}$)}
In this half-sphere, the angular part of the KG equation is
\begin{equation}
	\frac{1}{\sin\theta}\partial_{\theta}(\sin\theta \partial_{\theta}Y_{1}(\theta))+(\lambda_{1}-\frac{m^{2}+2(m q Q_{m}+q^{2}Q_{m}^{2})(1-\cos\theta)}{\sin^{2}\theta})Y_{1}(\theta)=0.
\end{equation}
Defining $\chi=\cos\theta$ and plugging it into the above angular equation, we can obtain the following differential equation
\begin{equation}
	\frac{d^2 Y_{1}}{d\chi^2}+(\frac{1}{\chi-1}+\frac{1}{\chi+1})\frac{dY_{1}}{d\chi}+[-\frac{m^{2} }{2(\chi-1)}+\frac{(2  q Q_{m}+m)^{2}}{2(\chi+1)}-\lambda_{1}]\frac{Y_{1}}{(\chi-1)(\chi+1)}=0.
	 \label{eqforboost}
\end{equation}
The above is a Fuchs-type equation with three singularities $(-1,1,\infty)$. The general solutions of it can be expressed by hypergeometric functions as
\begin{equation}
	\begin{aligned}
	Y_{1}(\chi)=&C_{1}  (1-\chi)^{\frac{-m}{2}}(1+\chi)^{\frac{ m +2 q Q_{m}}{2}} {}_2F_{1}(\alpha_{1},\beta_{1};\gamma_{1};z_{1})+\\
	&C_{2}(1-\chi)^{\frac{m}{2}}(1+\chi)^{\frac{ -m -2 q Q_{m}}{2}} {}_2F_{1}(\alpha_{2},\beta_{2};\gamma_{2};z_{2}),
		\end{aligned}
	\label{key25}
\end{equation}
where $C_{1},C_{2}$ are constants. The parameters in the hypergeometric functions are given by
\begin{equation}
\begin{aligned}
	\alpha_{1}&=\frac{1}{2}(1+2 q Q_{m}+\sqrt{1+4\lambda_{1}}), \quad  \alpha_{2}=\frac{1}{2}(1-2 q Q_{m}+\sqrt{1+4\lambda_{1}});\\
	\beta_{1}&=\frac{1}{2}(1+2 q Q_{m}-\sqrt{1+4\lambda_{1}}), \quad  \beta_{2}=\frac{1}{2}(1-2 q Q_{m}-\sqrt{1+4\lambda_{1}}); \\
	\gamma_{1}&=1-m,\quad \quad \quad \quad \quad \quad \quad \quad \qquad \gamma_{2}=1+m;\\
	z_{1}&=\frac{1}{2} (1-\chi), \quad \quad \quad \quad \quad \quad \quad \quad \quad z_{2}=\frac{1}{2} (1-\chi).
\end{aligned}
\end{equation}

In this case, the radial part of the KG equation is
\begin{equation}
	\Box\frac{d}{dr}(\Box\frac{dR}{dr})+U_{1}R=0,
\end{equation}
where
\begin{equation}
	U_{1}=(\omega r^2-q Q_e r)^2+\Box (q^2 Q^2_m-\mu^{2}r^{2}-\lambda_{1}).
	\label{key2}
\end{equation}

\subsection{The South Half-Sphere of the Black Hole( $\frac{\pi}{2}\leqslant\theta\leqslant\pi$)}

In this  half-sphere, the angular part of the KG equation is
\begin{equation}
	\frac{1}{\sin\theta}\partial_{\theta}(\sin\theta \partial_{\theta}Y_{2}(\theta))+(\lambda_{2}-\frac{m^{2}-2(m q Q_{m}-q^{2}Q_{m}^{2})(1+\cos\theta)}{\sin^{2}\theta})Y_{2}(\theta)=0.
\end{equation}
Similarly, we define $\eta=\cos\theta$ and the above equation can be rewritten as
\begin{equation}
	\frac{d^2 Y_{2}}{d\eta^2}+(\frac{1}{\eta-1}+\frac{1}{\eta+1})\frac{dY_{2}}{d\eta}+[\frac{m^{2} }{2(\eta+1)}-\frac{(2  q Q_{m}-m)^{2}}{2(\eta-1)}-\lambda_{2}]\frac{Y_{2}}{(\eta-1)(\eta+1)}=0.
	\label{key}
\end{equation}
The above is also a Fuchs-type equation with three singularities $(-1,1,\infty)$. The general solutions of it can be expressed by hypergeometric functions as
\begin{equation}
	\begin{aligned}
		Y_{2}(\eta)=&C_{3} (1-\eta)^{\frac{-m+2 q Q_{m}}{2}}(1+\eta)^{\frac{ m}{2}} {}_2F_{1}(\alpha_{3},\beta_{3};\gamma_{3};z_{3})-\\
		&C_{4}(1-\eta)^{\frac{m-2 q Q_{m}}{2}}(1+\eta)^{\frac{-m}{2}} {}_2F_{1}(\alpha_{4},\beta_{4};\gamma_{4};z_{4}),
	\end{aligned}
\label{key26}
\end{equation}
where $C_{3},C_{4}$ are  constants. The parameters in the hypergeometric functions are given by
\begin{equation}
	\begin{aligned}
		\alpha_{3}&=\frac{1}{2}(1+2 q Q_{m}+\sqrt{1+4\lambda_{2}}), \quad  \alpha_{4}=\frac{1}{2}(1-2 q Q_{m}+\sqrt{1+4\lambda_{2}});\\
		\beta_{3}&=\frac{1}{2}(1+2 q Q_{m}-\sqrt{1+4\lambda_{2}}), \quad  \beta_{4}=\frac{1}{2}(1-2 q Q_{m}-\sqrt{1+4\lambda_{2}}); \\
		\gamma_{3}&=1-m+2 q Q_{m},\quad \quad \quad \quad \quad \quad  \gamma_{4}=1+m-2 q Q_{m};\\
		z_{1}&=\frac{1}{2} (1-\eta), \quad \quad \quad \quad \quad \quad \quad \quad \quad z_{4}=\frac{1}{2} (1-\eta).
	\end{aligned}
\end{equation}

In this case, the radial part of the KG equation is given by
\begin{equation}
	\Box\frac{d}{dr}(\Box\frac{dR}{dr})+U_{2}R=0,
\end{equation}
where
\begin{equation}
	U_{2}=(\omega r^2-q Q_e r)^2+\Box (q^2 Q^2_m-\mu^{2}r^{2}-\lambda_{2}).
	\label{key3}
\end{equation}
\subsection{Analysis of The Angular Functions}
In order to ensure that the angular functions $Y_{i}(\theta)$ $(Y_{1}(\chi), Y_{2}(\eta))$ are finite at the north and south poles, the general solutions in equations \eqref{key25} and \eqref{key26} are chosen as follows,
 \begin{equation}
Y_{1}(\chi)=C_{2}(1-\chi)^{\frac{m}{2}}(1+\chi)^{\frac{ -m -2 q Q_{m}}{2}} {}_2F_{1}(\alpha_{2},\beta_{2};\gamma_{2};z_{2}),
 \end{equation}
\begin{equation}
	Y_{2}(\eta)=C_{3} (1-\eta)^{\frac{-m+2 q Q_{m}}{2}}(1+\eta)^{\frac{ m}{2}} {}_2F_{1}(\alpha_{3},\beta_{3};\gamma_{3};z_{3}).
\end{equation}
We also have the following remarks on the parameters in the angular functions,
\begin{enumerate}[I:]
	\item Finiteness of the factors $(1-\chi)^{\frac{m}{2}}$ and $(1+\eta)^{\frac{ m}{2}}$ implies
	\begin{equation}
m\geqslant0.
	\end{equation}
	\item  In order for the convergence of the  functions $_2F_1(\alpha_{2}$, $\beta_{2};\gamma_{2};z_{2})$, $_2F_1(\alpha_{3}$, $\beta_{3};\gamma_{3};z_{3})$ when $|z_2|,|z_{3}|\leqslant 1$, we obtain the charge quantization condition and constraints on $\lambda_{1},\lambda_{2}$
	\begin{equation}
	qQ_m= \text{integer}, \lambda_{1}=\lambda_{2}=l(l+1) ,\quad l>q Q_m .
	\label{key17}
	\end{equation}
\end{enumerate}

\subsection{The Radial Equation of Motion}
Now, we study the radial part of the equation of motion of the scalar perturbation. Based on the discussion of the angular  part of the equation of motion,  radial equations \eqref{key2} and \eqref{key3} can be rewritten as
\begin{equation}
	\Box\frac{d}{dr}(\Box\frac{dR}{dr})+UR=0,
	\label{key4}
\end{equation}
where
\begin{equation}
	U=(\omega r^2-q Q_e r)^2+\Box (q^2 Q^2_m-\mu^{2}r^{2}-\lambda).
\end{equation}

 In order to discuss the asymptotic solutions of the radial function near the outer horizon of the black hole, it is convenient to use the tortoise coordinate. Define the tortoise coordinate $r_{\star}$ by the  equation
\begin{equation}
\frac{dr_{\star}}{dr}=\frac{r^{2}}{\Box},
\end{equation}
and define a new radial function as $\xi=rR$, the radial equation\eqref{key4} can be rewritten as
\begin{equation}
\frac{d^{2}\xi}{dr^{2}_{\star}}+U_{1}\xi=0,
\end{equation}
where
\begin{equation}
U_{1}=\frac{U}{r^{4}}-\frac{ \Box}{r^{3}}\frac{d}{dr}(\frac{\Box}{r^{2}}).
\end{equation}

Here the suitable boundary conditions we need are  ingoing wave near the outer horizon ($r_{\star} \to -\infty$, i.e. $r\to r_{+}$) and exponentially decaying wave at spatial infinity($r_{\star}\to +\infty$, i.e. $r\to +\infty$). Hence, the radial equation of motion has the following asymptotic solutions
\begin{equation}
	\xi \thicksim\left\{ \begin{array}{l}
		{e^{-i(\omega-\omega_{c})r_{\star}}},\quad \,\,r_{\star}\to -\infty ,\\
		\\
		{e^{-\sqrt{\mu^2-\omega^2}r_{\star}}},\quad \,\,r_{\star}\to +\infty .\\
	\end{array} \right.
\end{equation}
In the above equation, the critical angular frequency $\omega_c$ is defined as
\begin{equation}
\omega_{c}=q \Psi,
\end{equation}
where  $\Psi$ is the electromagnetic potential of the outer horizon of the dyonic RN black hole, $\Psi=Q_e/r_+$. The superradiant condition for an electrically charged massive scalar perturbation on the dyonic RN black hole background is
\begin{equation}
\omega<\omega_{c}=\frac{q Q_{e}}{r_{+}}.
\label{key11}
\end{equation}
The bound state condition at spatial infinity for the scalar perturbation is
\begin{equation}
	 \omega^2<\mu^2.
	 \label{key12}
\end{equation}

\section{Effective Potential and Its Asymptotic Analysis}
 In this section, we will derive the effective potential from the radial equation of motion and analyze its asymptotic behaviours at the horizons and spatial infinity when the  parameters of the scalar field and the black hole satisfy the bound state condition $\omega^2<\mu^2$ and the superradiance condition $0<\omega<qQ_e/r_+$.

Define a new radial function $\psi$ by $\psi=\Box^\frac{1}{2}R$, then the radial equation of motion\eqref{key4} can be transformed into a Schrodinger-like equation
\begin{equation}
	\frac{d^2 \psi}{dr^2}+(\omega^2-V)\psi=0,
\end{equation}
where
\begin{equation}
	V=\omega^2-\frac{U+(r_{+}-M)^{2}}{\Box^2}
\end{equation}
is the effective potential experienced by the scalar perturbation field. If there is no potential well outside the outer horizon of the dyonic RN black hole,
the system composing of the charged massive scalar perturbation field and the dyonic RN black hole is superradiantly stable.

The asymptotic behaviors of the effective potential $V$ at the two horizons and spatial infinity are respectively
\begin{equation}\label{key13}
V(r\to r_-)\to -\infty, ~
	V(r\to r_+)\to -\infty;
\end{equation}
\begin{equation}
	V(r\to +\infty)\to \mu ^2+\frac{2}{r} f(\omega)+O\left(\frac{1}{r^2}\right).
\end{equation}
The asymptotic behavior of the derivative of V at spatial infinity is
\begin{equation}
	V^{'}(r\to +\infty)\to -\frac{ 2}{r^2} f(\omega)+O\left(\frac{1}{r^3}\right),
	\label{key14}
\end{equation}
where the function  $f(\omega)$ is
\begin{equation}
	f(\omega)=-2 M \omega^2+ q Q_e \omega+ M \mu^2.
\end{equation}

Now, let's prove that $f(\omega)$ is positive with the conditions \eqref{key11} and \eqref{key12}.   Obviously, there are one negative and one positive real roots  for $f(\omega)=0$  and the positive one is
\begin{equation}
	\omega_{+}=\frac{q Q_{e}+\sqrt{q^{2} Q_{e}^2+8M^2 \mu^2}}{4M}.
\end{equation}
In order to ensure $f(\omega)$ is positive, we just need to prove $0<\omega<\omega_{+}$. We will discuss this in two cases.

$\bullet$ Case I: $\omega<\mu<\frac{q Q_{e}}{r_{+}}$
\\
\begin{equation}
\omega_{+}=\frac{q Q_{e}+\sqrt{q^{2} Q_{e}^2+8M^2 \mu^2}}{4M}=\frac{q Q_{e}}{4 M}+\sqrt{\frac{q^{2} Q_{e}^{2}}{16 M^{2}}+\frac{\mu^{2}}{2}}.
\end{equation}
Since $r_{+}=M+\sqrt{M^{2}-Q_{e}^{2}-Q_{m}^{2}}>M$, we get an inequality as follows
\begin{equation}
\omega_{+}=\frac{q Q_{e}}{4 M}+\sqrt{\frac{q^{2} Q_{e}^{2}}{16 M^{2}}+\frac{\mu^{2}}{2}} > \frac{\mu}{4}+\sqrt{\frac{\mu^{2}}{16}+\frac{\mu^{2}}{2}}=\mu > \omega.
\end{equation}

$\bullet$ Case II: $\omega<\frac{q Q_{e}}{r_{+}}<\mu$\\ \\
 We can also easily get
\begin{equation}
	\omega_{+} >\frac{q Q_{e}}{4 r_{+}}+\sqrt{\frac{q^2 Q_{e}^2}{16 r_{+}^2}+\frac{q^2 Q_{e}^2}{2 r_{+}^2}}=\frac{q Q_{e}}{4 r_{+}}+\sqrt{\frac{ 9q^2 Q_{e}^2}{16 r_{+}^2}}=\frac{q Q_{e}}{r_{+}}>\omega.
\end{equation}

From the three equations of \eqref{key13}-\eqref{key14}, we know that there is at least one extreme between the inner horizon and the outer horizon ($r_{-}<r<r_{+}$) and there exists at least one maximum outside the outer horizon ($r>r_{+}$) since $f(\omega)>0$.  There is no potential well near the spatial infinity and the system may be superradiantly stable.
 In next section, we will go a step further and find the regime, satisfied by the parameters of the system, where there is only one maximum outside the outer horizon for the effective potential and no  potential well exists, which means the system is  superradiantly stable.

\section{Analysis of Superradiant Stability}
In this section, we will find the regions in the parameter space where the system of dyonic RN black hole and massive scalar perturbation is superradiantly stable. We determine the parameter regions by considering the extremes of the effective potential in the range $r_{-}<r<+\infty$.

\subsection{Explicit Expression of Derivative of $V$}
\quad \quad Now,  we define a new variable $y$, $y=r-r_{-}$. The expression of the derivative of the effective potential $V$ is
\begin{eqnarray}\nonumber
&&V'(r)=\frac{-2(Ar^4+Br^3+Cr^2+Dr+E)}{\bigtriangleup^{3}}\\
&=&V'(y)=\frac{-2 (A_{1}y^4+B_{1}y^3+C_{1}y^2+D_{1}y+E_{1})}{\bigtriangleup^{3}},
\end{eqnarray}
where
\begin{equation}
A_{1}=A;\quad B_{1}=(4r_{-})A_{1}+B,
\end{equation}
\begin{equation}
C_{1}=(6r_{-}^{2})A_{1}+(3r_{-})B_{1}+C,
\end{equation}
\begin{equation}
D_{1}=(4 r_{-}^{3})A_{1}+(3 r_{-}^{2})B_{1}+(2r_{-})C_{1},
\end{equation}
\begin{equation}
E_{1}=(r_{-}^{2})A_{1}+(r_{-}^{3})B_{1}+(r_{-}^{2})C_{1}+(r_{-})D_{1}+E.
\end{equation}
Explicitly,
\begin{flalign}
	\begin{split}
\qquad \quad A_{1}= - 2 M \omega^2 + q Q_e \omega+M \mu^2,
	\end{split}&
\end{flalign}
\begin{flalign}
	\begin{split}
		\qquad \quad	B_{1}=&-2\left(8 M^2-6 M r_++r_+^2\right)\omega ^2+2 q Q_e  \left(5 M-2 r_+\right)\omega\\
		&+\mu ^2 \left(6 M^2-6 M r_++r_+^2\right)-q^2 \left(Q_e^2+Q_m^2\right)+\lambda,
	\end{split}&
\end{flalign}
\begin{flalign}
	\begin{split}
\qquad \quad	C_{1}=&-6  \left(2 M-r_+\right){}^3 \omega ^2+9 q Q_e\left(2 M-r_+\right){}^2 \omega\\
	&+3((M-r_{+})(\mu^{2}(2M-r_{+})^{2}-q^{2}Q_{e}^{2}-q^{2}Q_{m}^{2}+\lambda)-Mq^{2}Q_{e}^{2}),
	\end{split}&
\end{flalign}
\begin{flalign}
	\begin{split}
\qquad \quad	D_{1}=& -2 \left(4 M-3 r_+\right) \left(2 M-r_+\right){}^3 \omega ^2 +2 q Q_e  \left(7 M-5 r_+\right) \left(2 M-r_+\right){}^2 \omega\\
	&+2 q^2 (-Q_m^2 (M^2-5 Q_e^2)+4 Q_e^4+Q_m^4)-2 q^2 Q_e^2 (3 M (r_+-2 M)\\
	&+2 \mu ^2 (2 M^2-3 M r_++r_+^2){}^2+2 (Q_e^2+Q_m^2))-12 M q^2 Q_e^2 r_-\\
	&+2\left(M-r_+\right){}^2(\lambda-1),
	\end{split}&
\end{flalign}
\begin{flalign}
	\begin{split}
\qquad \quad	E_{1}=\left(r_+-r_-\right) \left(q Q_e-\omega r_- \right){}^2r_-^2 +\frac{1}{4} \left(r_+-r_-\right)^3.
	\end{split}&
\end{flalign}
Because we are interested in the extremes of effective potential $V$, i.e. the roots of $V'$, in the next we mainly analyze the numerator of $V'$, which is
\begin{equation}
g(y)=A_{1}y^4+B_{1}y^3+C_{1}y^2+D_{1}y+E_{1}.
\end{equation}

\subsection{Analysis of Roots of $V'=0$}
From the asymptotic behaviors of the effective potential at the inner and outer horizons and spatial infinity, we know that there are at least two positive roots for
$g(y)=0$. These two positive real roots are denoted by $y_{1}$, $y_{2}$,  namely
 \begin{equation}
 y_{1}>0,\quad y_{2}>0.
 \end{equation}

The numerator of the derivative of the effective potential $g(y)$ is a quartic polynomial in $y$. The equation $g(y)=0$ has four roots, which are denoted by $y_{1},y_{2},y_{3}$ and $y_{4}$. By the Vieta theorem, we have

\begin{align}
y_{1}+y_{2}+y_{3}+y_{4}&=-\frac{B_{1}}{A_{1}},\\
y_{1}y_{2}+y_{1}y_{3}+y_{1}y_{4}+y_{2}y_{3}+y_{2}y_{4}+y_{3}y_{4}&=\frac{C_{1}}{A_{1}},\label{key16}\\
y_{1}y_{2}y_{3}+y_{1}y_{2}y_{4}+y_{1}y_{3}y_{4}+y_{2}y_{3}y_{4}&=-\frac{D_{1}}{A_{1}},\\
y_{1}y_{2}y_{3}y_{4}&=\frac{E_{1}}{A_{1}}.\label{key15}
\end{align}
It is worth an immediate remark here. The four roots are not necessary all real. $y_3, y_4$ may be complex roots and $y_3=y^*_4$.  We suppose the four roots are all real. This is the most worse case and we just want to find a sufficient condition for the system to be superradiantly stable.

In the following, we would like to analyze the three coefficients $A_{1}, C_{1}$ and $E_{1}$.
The coefficient $A_{1}$ has already proved to be positive in previous section.
For coefficient $E_{1}$, it is easy to see that
\begin{equation}
r_{+}>r_{-},\quad 	E_{1}=\left(r_+-r_-\right) \left(q Q_e-\omega r_- \right){}^2r_-^2 +\frac{1}{4} \left(r_+-r_-\right)^3>0.
\end{equation}
Thus, according to the equation \eqref{key15} of Vieta theorem, there are two cases for the four roots of the equation $g(y)=0$: two positive roots and two negative roots or all four positive roots. If  $C_{1}$ is less than 0, these four roots are two positive and two negative roots, there is no potential well outside the outer horizon and the system is superradiantly stable.

In the following, we will find the regime where $C_{1}<0$. $C_{1}$ can be treated as a quadratic polynomial in $\omega$,
\begin{equation}
	\begin{aligned}
			C_1(\omega)=&-6  \left(2 M-r_+\right){}^3 \omega ^2+9 q Q_e\left(2 M-r_+\right){}^2 \omega\\
		&+3((M-r_{+})(\mu^{2}(2M-r_{+})^{2}-q^{2}Q_{e}^{2}-q^{2}Q_{m}^{2}+\lambda)-Mq^{2}Q_{e}^{2}).
	\end{aligned}
\end{equation}
Given the condition  $(M-r_{+})<0$ and the inequality \eqref{key17}, $-q^{2}Q_{m}^{2}+\lambda>0$, we have
\begin{equation}
	C_1(\omega)<-6  r_-^3  \omega ^2+9 q Q_e r_-^2 \omega  +3(M-r_{+})(\mu^{2}r_{-}^{2}-q^{2}Q_{e}^{2})-3Mq^{2}Q_{e}^{2}.
\end{equation}
Using the equation $M-r_+=r_--M$, the above inequality can be rewritten as
\begin{equation}
	C_1(\omega)<-6  r_-^3  \omega ^2+9 q Q_e r_-^2 \omega  +\frac{3}{2}\mu^{2} r_-^{3}-\frac{3}{2}\mu^{2}r_{+}r_{-}^{2}-3 q^{2}Q_{e}^{2}r_{-}.
\end{equation}
Considering the bound state condition\eqref{key12} and the above inequality, we have
\begin{equation}
C_1(\omega)<-6  r_-^3  \omega ^2+9 q Q_e r_-^2 \omega  +\frac{3}{2}\omega^{2} r_-^{3}-\frac{3}{2}\omega^{2}r_{+}r_{-}^{2}-3 q^{2}Q_{e}^{2}r_{-}\equiv 3r_- h(\omega),
\end{equation}
where
\begin{equation}
h(\omega)=-\frac{1}{2} (3 r_-^2+r_+ r_-) \omega ^2 +3 q Q_e r_- \omega  -q^2 Q_e^2.
\end{equation}

\begin{figure}[b]
	\centering
	\includegraphics[width=0.7\linewidth]{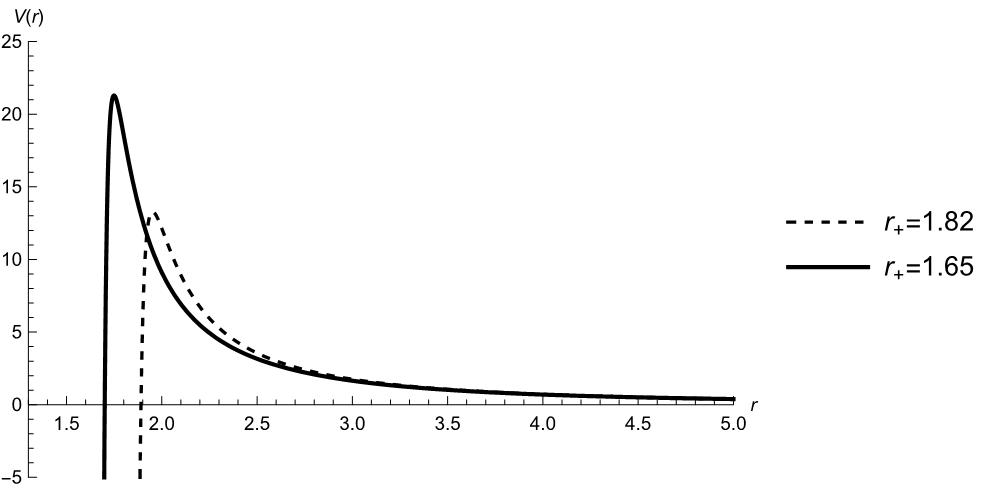}
	\caption{Superradiantly stable effective potentials. The black hole mass are chosen as $M=1$. The charges are chosen as $Q_{e}=0.4, Q_{m}=0.4$ and $Q_{e}=0.7,Q_{m}=0.3$ for the solid curve and dashed curve respectively. The other parameters are chosen as $l=2,\omega=0.02, \mu=0.05, q=0.1 .$}
	\label{fig1}
\end{figure}

When $h(\omega)<0$, we have $C_1<0$. Regarding $h(\omega)$ as a quadratic function of $\omega$, the discriminant of $h$, $\Delta_{h}$, is
\begin{equation}
	\Delta_{h}=q^2  Q_e^2 r_{-}(3r_- -2r_+).
\end{equation}
It is obvious that when
\begin{equation}\label{f-r}
	\frac{r_{-}}{r_{+}}<\frac{2}{3},
\end{equation}
$\Delta_{h}<0$, $h(\omega)<0$ and $C_1(\omega)<0$.
Two examples of the superradiantly stable effective potential are illustrated in Fig.\eqref{fig1}

\section{Summary}
We have investigated  the superradiant stability property of a system consisting of a dyonic RN black hole and an electrically charged massive scalar perturbation. The dyonic RN black hole is an electrically and magnetically charged black hole which is also spherical and asymptotically flat.
 The equation of motion of the scalar perturbation in the dyonic RN black hole background is separated into angular and radial parts. We discuss
  the angular equations in two cases and obtain the charge quantization condition
  \begin{equation*}
qQ_m=\text{integer},
\end{equation*}
and a bound on the magnetic charge of the black hole
\begin{equation*}
l(l+1)>q^{2}Q^{2}_{m}.
\end{equation*}
The radial equation of motion is transformed into a Schrodinger-like equation and the effective potential $V$ experienced by the scalar perturbation is derived. Through the analysis of the effective potential, we find the following simple regime
\begin{equation*}
 \frac{r_{-}}{r_{+}}<\frac{2}{3},
\end{equation*}
where there exists no trapping potential outside the black hole horizon and the system is superradiantly stable. Compared with the purely electrically charged RN black hole case, our result implies that  the magnetic filed  makes the black hole and scalar perturbation system more superradiantly unstable.

{\textbf{Acknowledgements:\\}}
This work is partially supported by Guangdong Major Project of Basic and Applied Basic Research (No. 2020B0301030008), Science and Technology Program of Guangzhou (No. 2019050001) and Natural Science Foundation of Guangdong Province (No. 2020A1515010388, No. 2020A1515010794).

\end{document}